# Spatial correlations in geographical spreading of COVID-19 in USA

Troy McMahon[1], Adrian Chan[2], Shlomo Havlin[2], and Lazaros K. Gallos[1]

[1] DIMACS, Rutgers University, Piscataway, New Jersey 08854, USA.
[2] Department of Physics, Bar-Ilan University, Ramat Gan 52900, Israel.

**Abstract**

The global spread of the COVID-19 pandemic has followed complex pathways, largely attributed to the high virus infectivity, human travel patterns, and the implementation of multiple mitigation measures. The resulting geographic patterns describe the evolution of the epidemic and can indicate areas which are at risk of an outbreak. Here, we analyze the spatial correlations of new active cases in USA at the county level and characterize the extent of these correlations at different times. We show that the epidemic did not progress uniformly and we identify various stages which are distinguished by significant differences in the correlation length. Our results indicate that the correlation length may be large even during periods when the number of cases declines. We find that correlations between urban centers were much more significant than between rural areas and this finding indicates that long-range spreading was mainly facilitated by travel between cities, especially at the first months of the epidemic. We also show the existence of a percolation transition in November 2020, when the largest part of the country was connected to a spanning cluster, and a smaller-scale transition in January 2021, with both times corresponding to the peak of the epidemic in the country.

# INTRODUCTION

The high infection rate of COVID-19 [1], combined with modern society behavioral patterns and a high volume of travel [2], enabled the rapid spreading of the virus globally [3]. The COVID-19 pandemic is among the few examples of an infectious disease which spread widely in a relatively short amount of time and whose evolution was closely monitored and documented [4,5]. In the absence of restriction measures, the entire population would have been exposed to the virus resulting to a much higher number of active cases and consequently deaths. Many measures have been implemented at different countries and most of them target to limiting contacts between large number of people [6] and limiting travel over large distances [7]. These measures include travel restrictions, quarantine, social distancing, lockdowns, curfews, and others [8]. Such restrictions have led to a varying degree of success which is difficult to estimate since individuals may not have followed state directives or the measures may have been inadequate [9]. However, it is clear that these mitigation measures have had a significant impact on the evolution of the pandemic and that the geographical spreading would have been very different if the virus was left to spread uncontrolled within the population [10].

The special features of this virus and the unprecedented global response present potentially novel paths of disease transmission that have not been observed in modern times [11]. The combination of these transmission paths manifests itself as the number of new active cases over a region of interest, e.g. a county in USA, and these numbers are reported daily. The geographical arrangement of variables such as the number of new cases gives rise to larger scale spatial patterns which span broader areas on the map. The analysis of these patterns and their pertinent features can provide important information on the extent of the epidemic at a given time and areas which may be at higher risk of an outbreak. In practice, it is possible to identify geographical clusters whose connectivity is based on similar local levels of infections or similar trends in the local progress of infection [12]. We can then assess how these clusters evolve with time, e.g. in terms of their size and persistence.

Here, we suggest that the use of spatial correlation statistics and cluster analysis of spreading indicators can provide valuable information on the extent of spatial spreading and how spatial correlations arose within and between geographical areas. Similar approaches have been shown to be very successful in other contexts of spreading [13,14], where they have provided important results and insight in problems related to spatial epidemics.

We find that during the one year of spreading from February 2020 to February 2021, when vaccinations started becoming widely available, there were three main phases in terms of spatial spreading, which can be roughly described as localized, dormant, and system-wide outbreak. Interestingly, if we consider the whole country to represent one system, then these three phases can be compared to the progression of a disease in an individual, moving from an acute infection to false recovery to severe illness. In spring 2020 (the localized, acute phase), spreading was contained within small clusters and there were only a few local outbreaks mainly located in the Northeast. From May 2020 to October 2020, (the incubation phase) correlations in new active cases were weaker across the country but at the same time the underlying clusters started growing in size while still remaining mostly localized in space. November 2020 marks the beginning of the third stage (the system-wide outbreak phase), when correlations spanned the largest part of the country as these clusters merged. Even though this spanning cluster dissolved in less than a month, correlations remained strong for the rest of this time interval indicating that virus transmission could still increase at a fast pace.

It is also noteworthy that correlations among neighboring urban centers remained strong throughout this year. On the contrary, until November 2020 there were only weak or no correlations between rural areas, even for counties which are geographically close to each other. After November 2020, these correlations increased in strength and became comparable to those between urban counties. This behavior can also explain the percolation transition to a cluster spanning the largest part of the country at that time.

## METHOD

We analyze COVID-related data for USA at the county level using the Johns Hopkins dataset [5, 15]. The main quantity we study is the number of new COVID cases. We use a 7-day window in order to alleviate problems with inconsistent data reporting, such as weekend vs weekday testing patterns. Starting on February 1 2020, we aggregate the total number of newly infected cases in a given county over the past 7 days (including the given day) and calculate the daily average during this time period. We then convert this number to the average daily fraction of the population in each county that was infected during this week by dividing with the county population. In short, if $z_t(i)$ denotes the number of new cases in county $i$ on day $t$, then for week $T$ we calculate the fraction $Z_T(i)$ as:

$$Z_T(i) \equiv \frac{1}{7\,p_i} \sum_{t=7T-6}^{7T} z_t(i) \qquad (1)$$

where $p_i$ is the county population. We remove 697 counties with population less than 10000 because a small change in the number of cases in a small population can lead to large fluctuations, which results to a total of 2411 counties in our calculations. In this way, we create weekly maps for the infection rates of each county $Z_t(i)$ for the time period from February 1, 2020 to February 1, 2021. Some of the resulting maps are shown in Figure 1a. As expected, these maps indicate that the spatial coverage of the virus is not uniform but incidents are geographically clustered. These clusters change significantly with time, both in terms of their size and location. Our main goal here is to quantify these clusters through a spatial correlation analysis [16], so that we can detect the evolution of spreading and the impact of restriction measures.

We start by detrending the data for the correlation analysis. We use the differences method [17] where we consider the difference of $Z$ between two consecutive weeks T and T-1, i.e. $X_T(i)=\Delta Z_T(i)= Z_T(i)-Z_{T-1}(i)$. In practice, $X_T(i)$ measures the extent to which the relative number of cases increased or decreased on a given week compared to the previous week. We consider the equal-time two-point correlation function on week $T$, $C_T(r)$, which is the average of the correlation of $X_T$ over all counties at distance $r$, as measured between the centroids of two counties:

$$C_T(r) \equiv \frac{1}{|N(r)|} \frac{\sum_{N(r)} X_T(s)X_T(s+r) - m_{T,s} m_{T,s+r}}{\sqrt{\sigma_{T,s}^2 \sigma_{T,s+r}^2}}, \qquad (2)$$

where the average values *m* and variances $\sigma$ are defined below. There is no explicit dependence on time in this calculation and we calculate the spatial correlation using information only of a given snapshot in time and subsequently analyze these correlations at different times. For this calculation we consider the vector from a site **s** (as represented by the geographical coordinates of a county centroid) to all the centroids lying at a circle **s**+**r** of radius *r*=|**r**|. Because of the natural inhomogeneity in the geographical distribution of the counties, in practice there are not any counties at an exact distance *r* and we use all data values within the neighborhood of **r**. Therefore, in the above formula *N(r)* represents the set of pairwise distances between counties which are within a distance [r-dr,r+dr] from each other. In practice, we bin the distances logarithmically, so that dr increases by a factor 1.05 between successive bins after fixing the first bin in the range [0km,75km]. The average values $m_{T,s} = \sum_{N(r)} X_T(s)/|N(r)|$ and $m_{T,s+r} = \sum_{N(r)} X_T(s+r)/|N(r)|$ are calculated at the origin and the end of the vectors respectively. Similarly, the variances of these distributions are given by $\sigma^2_{T,s} = \sum_{N(r)} (X_T(s) - m_{T,s})^2/|N(r)|$ and $\sigma^2_{T,s} = \sum_{N(r)} (X_T(s+r) - m_{T,s+r})^2/|N(r)|$. The set of points *s* includes all counties in the area that we study. For example, in country-wide calculations **s** includes all the counties and since each county is counted as both an origin and an end of the vector, the two distributions $X_T(s)$ and $X_T(s+r)$ are identical. When we focus on a smaller area, such as a region or a state, then the set *s* includes only counties within this region while the set *s+r* includes points which may lie outside this region and therefore these two distributions are not necessarily the same.

The function $C_T(r)$ in Eq. (2) is an indicator of how correlation decays with distance on week *T* as we move away from a given point in space [18]. The correlation length, ξ, is then defined as the minimum distance where this function assumes a value of 0, i.e. $C_T(\xi)=0$ [19]. Long-range correlations are manifested by a large correlation length, while ξ vanishes for a random distribution. In the context of an epidemic process, long-range correlations are a hallmark of virus transmission through travel between distant places [20]. If travel was severely limited and people could only interact locally then the evolution of the disease at different areas would be largely independent of each other which would be manifested by weak correlations and small correlation lengths.

## RESULTS

The evolution of the COVID epidemic has been highly inhomogeneous and did not follow spatial and temporal patterns of typical infectious diseases [21]. The implementation of quarantine and other travel restriction measures at relatively early stages of the outbreak resulted in the general absence of infected clusters persistently spanning a significant fraction of the country [22]. To study the extent of these areas we calculate the correlation function C(*r*) as a function of the distance *r* at different times, where the set **s** in Eq. (2) includes the 2411 counties with population greater than 10000 people. As expected, correlation values start at C(0)=1 and decay to 0, which also defines the correlation length ξ as the shortest distance where C(*r*)=0. In Fig. 2a we calculate C(*r*) for the first week of each month from March 2020 to February 2021. It is obvious that correlation varies significantly over time, both in terms of the local correlation strength at small distances (typically less than 100km) and in terms of the correlation length. In Fig. 2b we focus on some of the strongest correlations and use double-logarithmic axes to highlight their different behavior. For example, during July 2020 the local correlations are weak but persist over a distance of 800km, while in April 2020 the local correlations were much stronger and decayed faster with a correlation length of 400km.

We found that the epidemic patterns changed significantly over time, as shown in Fig. 2c where we isolate four C(r) curves per plot for clarity. In the first phase of the epidemic, from March to June 2020, correlations fluctuated around zero with the striking exception of April when local correlations were strong and decayed fast. Local correlations were higher in the next time interval, from July to October 2020, but remained relatively weak. However, the correlation length increased significantly during this time. While the relatively low infection numbers during summer seemingly suggest that the country is "recovering" from the pandemic, the increasing correlation length tells us the opposite - that the virus is silently taking root everywhere, which will lead to the eventual outbreak in the next phase. This indicates that the correlation length is not just a different way to look at the raw numbers but can reveal underlying phenomena that can serve as warning signs to policymakers not to relax restrictions too early. Indeed, during the next phase of the epidemic, from November 2020 to February 2021, the local correlations increased significantly and the correlation length remained relatively large. In a fast-spreading epidemic, the expectation is that correlations become strong and extend over long distances, especially if they are facilitated by long-range travel [23]. Here, we see that after an initial peak in April 2020, it took many months for correlations to increase and to remain strong over a long period of time.

Importantly, these curves are averages over all US counties but the evolution of the epidemic may be different at different regions of the country. In Fig. 2d we calculate correlations in each of the four Census Bureau-designated regions, i.e. Northeast, Midwest, South, and West [24]. In the examples shown, local correlations and correlation lengths vary quite significantly in different regions and while Northeast dominates in May, the West exhibits much stronger correlations in August.

The evolution of the C(r) curves carries a lot of information and we need some way to probe this information and compare different instances in time. If the functional form remained the same throughout this process then we could compare any parameters that would appear in a model that could describe the evolution (for example, if these curves could be described by a power-law we could compare the power-law exponents). As can be seen in the various plots of Fig. 2, the dependence of C(r) on r cannot be described by the same form. For example, in Fig. 2b some curves can be described by a modified power-law while others are fitted better by an exponential form, but there is not a uniform description over the entire time interval. Therefore, we choose to characterize this behavior by comparing both the strength of 'local' correlations, i.e. the value of C(r) at small distances around 50km, and the spatial extent of correlations through the value of the correlation length. In a sense, these two parameters capture the basic trends that we are interested in these plots: all curves decay from a given value at C(r<50km) down to C($\xi$)=0. Even though we lose the exact form of the decay, these two points define in very broad terms the extent and strength of the correlation function and here we are not concerned about the curvature of the line in the intermediate regime. The combination of these two parameters can already inform on whether counties tend to belong to large clusters and whether the influence within these clusters is strong or weak.

So, how do we characterize the COVID evolution to determine if spreading expands geographically or if it shrinks? As discussed above, the correlation length itself is important but is not fully adequate. Even if ξ is large, it is possible that correlations in shorter distances could be weak and, as a result, less influential. In Fig. 3a we plot the correlation function continuously from Feb 1, 2020 to Feb 1, 2021 in weekly increments. The correlation length is marked as the distance where C(*r*) becomes zero on each week. The correlation length fluctuates significantly but we can distinguish two peaks in April and July, followed by a rather constant high value from October to January. Time periods with stronger local correlations are identified by red color. Even though many of the correlation values may seem small, these are averages over the entire country where spreading may be very inhomogeneous.

The consideration of these two parameters allows us to plot the 'trajectory' of the epidemic in a phase space of C(r<50km) and ξ (Fig. 3b), where we plot these two quantities continuously for different points in time. To conceptualize this plot we can split the phase space into four quadrants. When C(r) is small, i.e. at the bottom quadrants, then correlations are weak. The large correlation lengths found at the quadrants at the right, indicate that epidemic can spread easily across large distances. From the diagram in Fig. 3a we can see that in April 2020 there was a significant increase in local correlation strength but correlation length was still relatively small. The trajectory returned close to the origin (indicating a random distribution of cases) until the end of summer, when both the intensity and length increased significantly. The trajectory has remained at the area of the upper right quadrant from October 2020 until the end of the year, which indicates extended correlations over space. Starting in January 2021, both the correlation length and the correlation strength decreased considerably, with an abrupt increase of C(r<50km) at the beginning of February 2021.

These trajectories can shed light on the different evolution of spreading in different areas of the country (Fig. 3c). For example, the correlation length in Northeast has remained relatively short (this region is also the smallest) but the local correlation strength increased significantly in April 2020 and January 2021. This shows that at the early stages of the epidemic there was a strong cluster in this area but it remained localized. In contrast to that, the correlation length in the West was consistently large since June 2020, with spikes of the intensity between September 2020 and January 2021. The behavior in the Midwest was also different. The trajectory there remained close to the origin from the beginning until October 2020 when both the intensity and correlation length increased. Notice that in all regions the correlation length in November 2021 grew larger, which is an indication that local clusters joined into a larger cluster spanning the majority of the country.

In typical epidemic processes, it is expected that correlations are higher in areas geographically close to each other, since the main method of transmission is close contact between individuals [25]. Travel represents another important mechanism which contributes to long-range correlations, where now the underlying mechanism is the direct transfer of the virus over a large distance through air or ground travel [26]. To determine the contribution of travel we calculated the correlation as a function of the distance in the case of urban areas vs rural areas. Here, we set an arbitrary criterion for an urban county as one with population larger than 250,000 people and a rural county with population less than that. Using this threshold, there are 273 urban and 2138 rural counties. We then calculate the correlation function for the new active cases between urban counties only, between rural counties, or between rural and urban. In Fig. 4a, we plot C($r$) for these three cases for the week of May 1, 2020. There are practically no correlations between rural areas even for short distances and there is only weak correlation between rural and urban areas, and this is true only for short distances. In contrast, correlations between urban places are much higher for distances up to roughly 300km. This is an important observation because at that time air travel was severely limited and passengers were heavily screened, while car transportation was not restricted. Therefore, there are two possible explanations. Assuming that the first significant center of the epidemic was located in the urban NYC area, either virus transmission was facilitated through car transportation from city to city or the city lifestyle made the virus spread easier in an urban environment and cities presented similar behavior independently of each other. Interestingly, in Fig. 4a correlations between urban places that are farther than 300km become largely uncorrelated, which favors the idea of local transmission through ground transportation.

Similar results were obtained at other times, such as in August 2020 shown in Fig. 4b, when the difference between cities and city-rural areas was more pronounced. Correlations between urban counties remained stronger. However, travel had increased during summer and the restriction measures were relaxed, which in the plot is manifested by smaller peaks at 300km and 600km. In January 2021 (Fig. 4c), the differences vanished and correlations were practically independent on the urban character of the counties. At that time, the correlation length increased to 800km indicating again a global outbreak covering the largest part of the country.

To explore the time evolution of these observations we calculated the average correlation <C($r$)> for the three cases (urban-urban, urban-rural, and rural-rural) over three distance intervals, $r$<250km (Fig. 4d), 250km<$r$<1000km (Fig. 4e), and $r$>1000km (Fig. 4f). When we consider short distances the correlation between urban areas is consistently higher than in the other two cases. In fact, from February to October 2020 the spreading in rural areas was uncorrelated when rural areas were involved which shows that the epidemic was largely contained in urban environments. After October 2020, correlations became stronger in all cases for short distances, and the behavior was consistent independently of the rural or urban character of the area.

In the case of intermediate distances (Fig. 4e) the value of <C($r$)> mostly fluctuates around 0 with a small peak of inter-city correlations from June to August 2020. This plot supports the idea of ground transportation transmission since long road trips are much more rare than local road trips and in these distances air travel is typically the preferred mode of transportation. For distances longer than 1000km, the inter-city correlations are consistently zero or weakly anti-correlated. This shows that not only spreading in the west coast was different than in the east coast but in general they had opposite trends (notice that these are averages over the entire country and it is possible that locally correlations may be much stronger or weaker than these averages).

As shown above, in November the average correlation length for the entire country was of the order of 800-1000km which is comparable to the size of the system. For reference, a 1000km-radius circle centered in the middle of the continental US would cover roughly half of the country (the distance from New York to Los Angeles is around 4000 km). This is a strong indication of a widespread epidemic which approaches the percolation threshold.

To determine how close the epidemic came to percolating throughout the country and when this happened we performed a clustering analysis [27]. For a given point in time, we consider all clusters created by connected counties whose weekly difference $\Delta Z_T(i)$ exceeds a given threshold. We create two types of clusters, depending on whether these differences are positive or negative. Positive clusters are areas where the epidemic has increased significantly over the past week and negative clusters are areas where it decreased. We determine the size of a cluster by the total geographic area covered by the counties that comprise this cluster. Fig. 5a shows that the largest clusters were relatively small and localized until September 2020, which was followed by a rapid increase in the size of the positive cluster which within two months covered an area of around $4.5 \times 10^6$ km$^2$ (the total area of the 48 contiguous states is roughly $8 \times 10^6$ km$^2$). This cluster was dissolved within two weeks and its size remained small with an exception of a smaller peak during January 2021. The size of the negative cluster remained small until after the large positive cluster was formed in November 2020. It is interesting that the two processes of positive or negative change above a given threshold are not completely synchronized and cannot fully explain each other, i.e. the extent and location of the negative clusters do not necessarily follow the positive cluster. This can be seen in the largest decrease of the positive cluster size compared to the smallest increase of the negative cluster in December 2020, as well as the second peak of the negative cluster in January 2021 which was not preceded by a positive cluster of comparable size.

In the maps of Fig. 5b we present the two largest positive and the two largest negative clusters at different times. In agreement with previous observations the clusters from April to October 2020 are relatively small. In October 2020, the clusters started getting bigger to the west and south of the country and in November 2020 there was a transition to a country-spanning positive rate cluster. This cluster started dissolving in December 2020, when the middle part of the country was connected through a negative rate cluster leaving connected areas of positive rates at the east and west parts of the country.

Clustering can also be used to identify areas of the country where the epidemic persisted the longest time as part of a large cluster. For this, we considered all the counties which belonged to a cluster of at least 10000 km$^2$ for a minimum of 1.5 months during a three-month period. The results in Fig. 5c show that from April to August 2020 there were only few isolated counties involved in the largest clusters. Contrary to that, a large part at the north of the country was consistently included in large clusters during the period of September to November 2020. For the next time period, cases were consistently increasing throughout the southern part of the country and mainly in the southwest. In general, we see that the epidemic has been spreading quickly but has not persisted over extended areas for more than a few months, perhaps as a result of implementing restriction measures when active cases were increasing at a local level.

## DISCUSSION

One loose interpretation of a correlation-based spatial cluster is that the virus may be transmitted more easily within counties in this cluster, compared to counties whose active cases are uncorrelated [28]. This is a phenomenological manifestation of the underlying transmission mechanisms, even though these mechanisms are not explicitly known [29]. This macroscopic evaluation of the epidemic footprint can identify areas of a country where spreading is highly correlated and since spatial correlations can exist almost independently of the current level of the number of active cases, they can potentially be used as warning signals.

Using spatial statistical analysis, our results indicate that the COVID-19 epidemic in continental USA went through different phases. The first localized clusters started dissolving quickly, which indicates that there was not any significant long-distance transmission during spring 2020. By the summer of 2020, many local clusters started emerging whose size was continuously increasing and by November 2020 they had merged into a country-spanning cluster. This formation was short-lived and even though local correlations remained strong, the global correlation length started decreasing.

A similar approach can be applied to smaller areas, such as regions or states. We have created an online tool, where the user can select individual states for further analysis. This analysis could provide additional information on how the epidemic evolved at a smaller geographic scale and a possible extension of this work is to relate differences in virus spreading with state-level mitigation measures.

## ACKNOWLEDGEMENTS

This work was supported by a joint NSF-BSF grant. TM and LKG were supported by NSF through DEB-2035297. AC and SH were supported by BSF through grant 2020645.

## ADDITIONAL INFORMATION

The webpage http://dimacs.rutgers.edu/~lgallos/COVID hosts an interactive tool which allows the reconstruction of the maps included in the paper at different points in time. The tool also displays plots and maps for individual states.

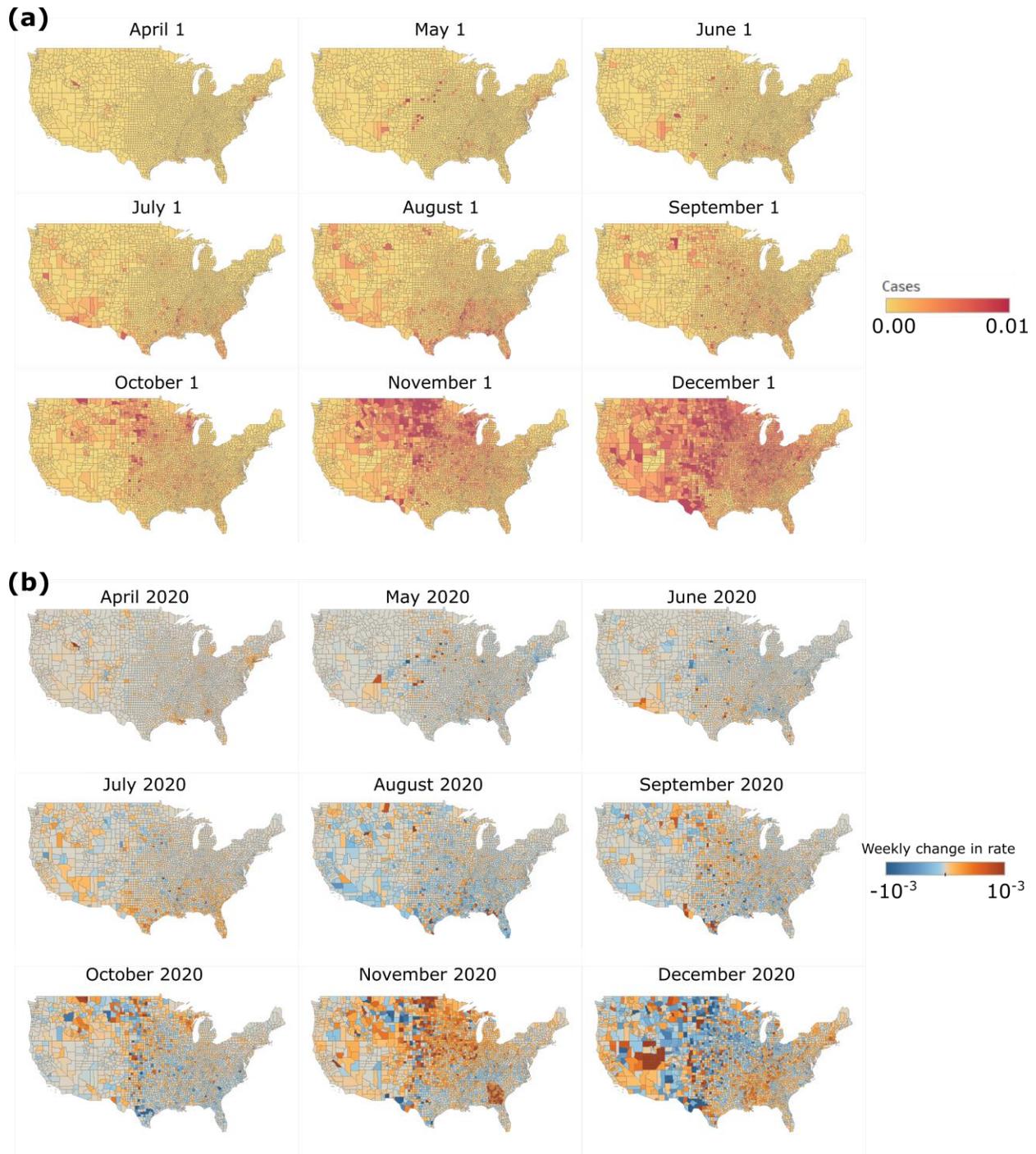

**FIGURE 1 | Evolution of COVID-19 spreading in the continental US. (a)** Average daily rate of new infections from April to December 2020 for the first week of the month at the county level. **(b)** Maps of the quantity $X_T(i)$, which corresponds to the change in the daily rate of infections between two consecutive weeks in the beginning of each month. Red color indicates an increase compared to the previous week and blue color indicates that the rate decreased.

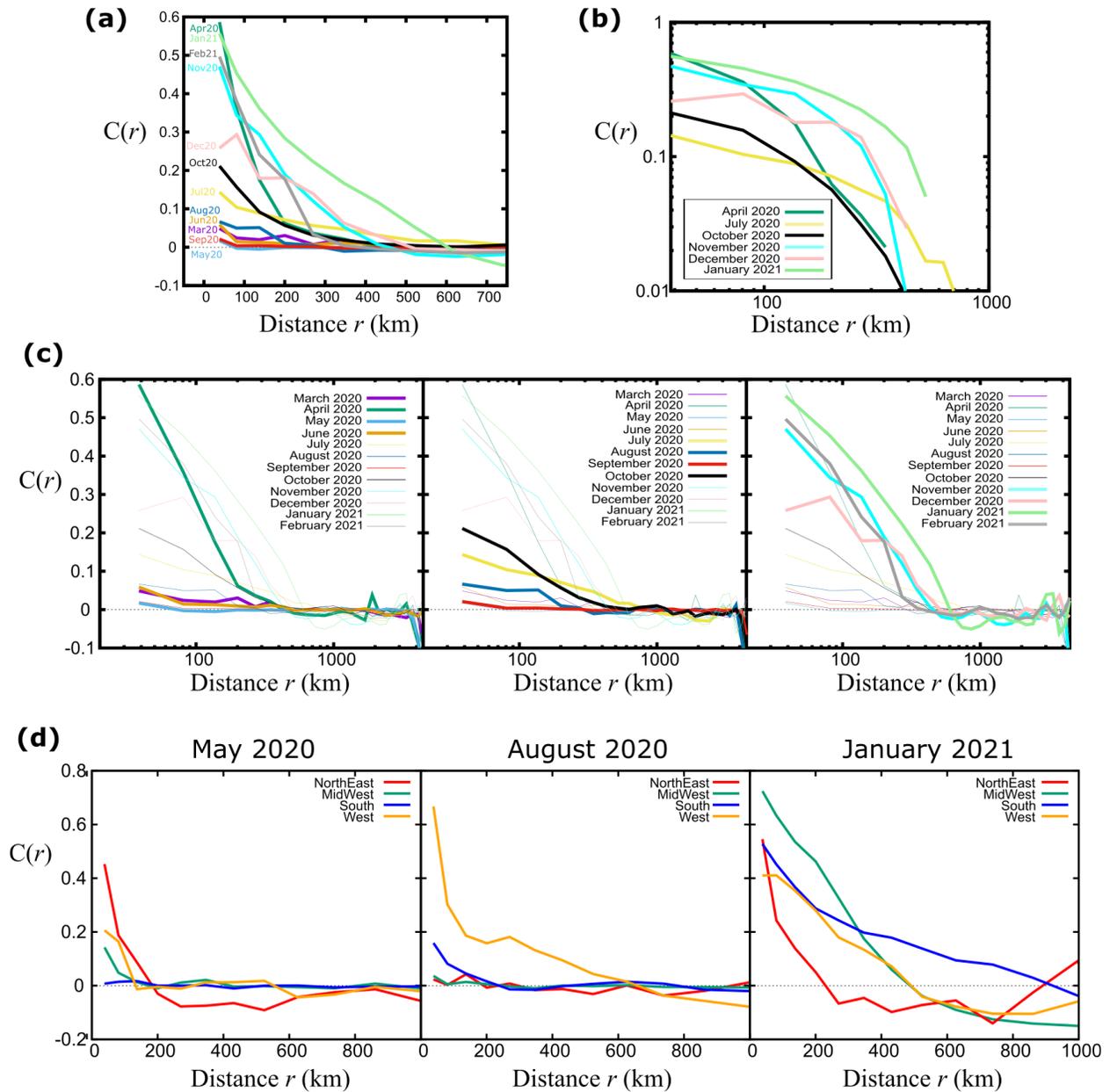

**FIGURE 2 | Decay of the correlation function with distance.** **(a)** The correlation function as a function of the distance, averaged over all counties in the United States. Different curves correspond to different instances of time, for the first two weeks of the month from March 2020 to February 2020. **(b)** Same plot as in (a) in double logarithmic axes for select months. **(c)** For clarity, we split the plot in (a) into three different time intervals and focus on four months within these intervals. **(d)** Comparison of the correlation functions, averaged over all counties within the four US regions, for May 2020, August 2020, and January 2021.

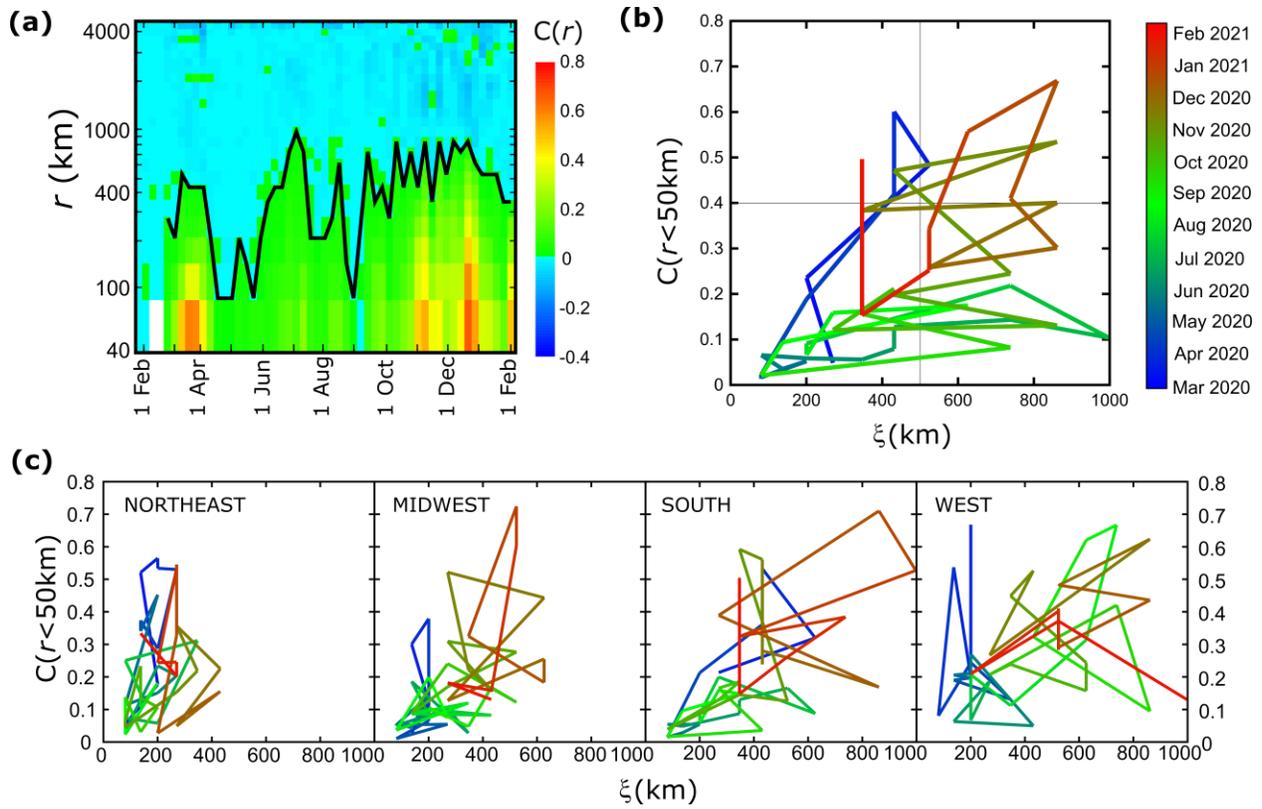

**FIGURE 3 | Time evolution of the correlation function.** (a) The value of the correlation function is shown as a function of the date (x-axis) and distance (y-axis). Red and green colors correspond to positive values while blue colors indicate negative values. The line that separates the two regimes describes the correlation length as a function of time. (b) The 'trajectory' of the epidemic is plotted by the values of local correlations C(r<50km) vs the correlation length, $\xi$, at different times for the entire country. (c) Same trajectory plots as in (b) for the four regions of US.

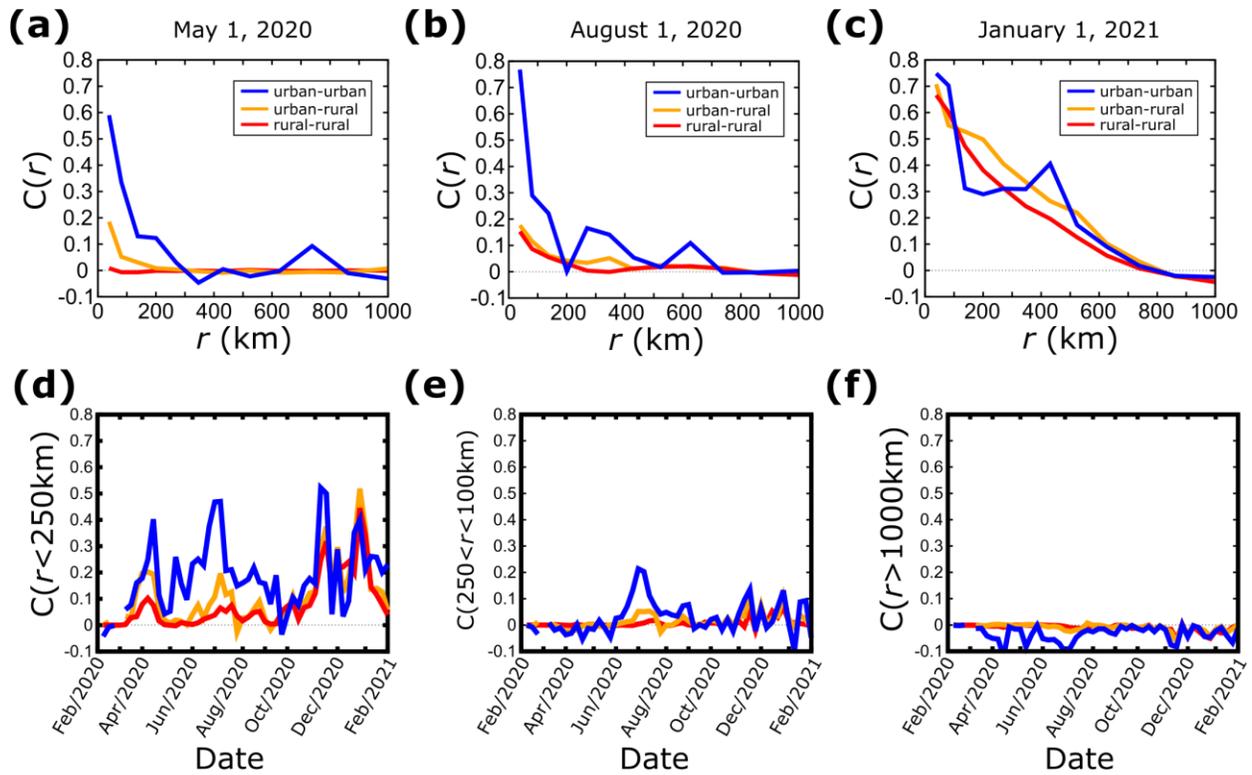

**FIGURE 4 | Spatial correlations between areas of different population (a-c)** Correlation function C(r) between urban-urban, urban-rural, and rural-rural counties, as a function of the distance at three different times. **(d)** Average correlation function between urban-urban, urban-rural, and rural-rural counties which are within 250km from each other, as a function of time. **(e)** Same plot as in (d) for county distances between 250km and 1000km. **(f)** Same plot as in (d) for distances longer than 1000km.

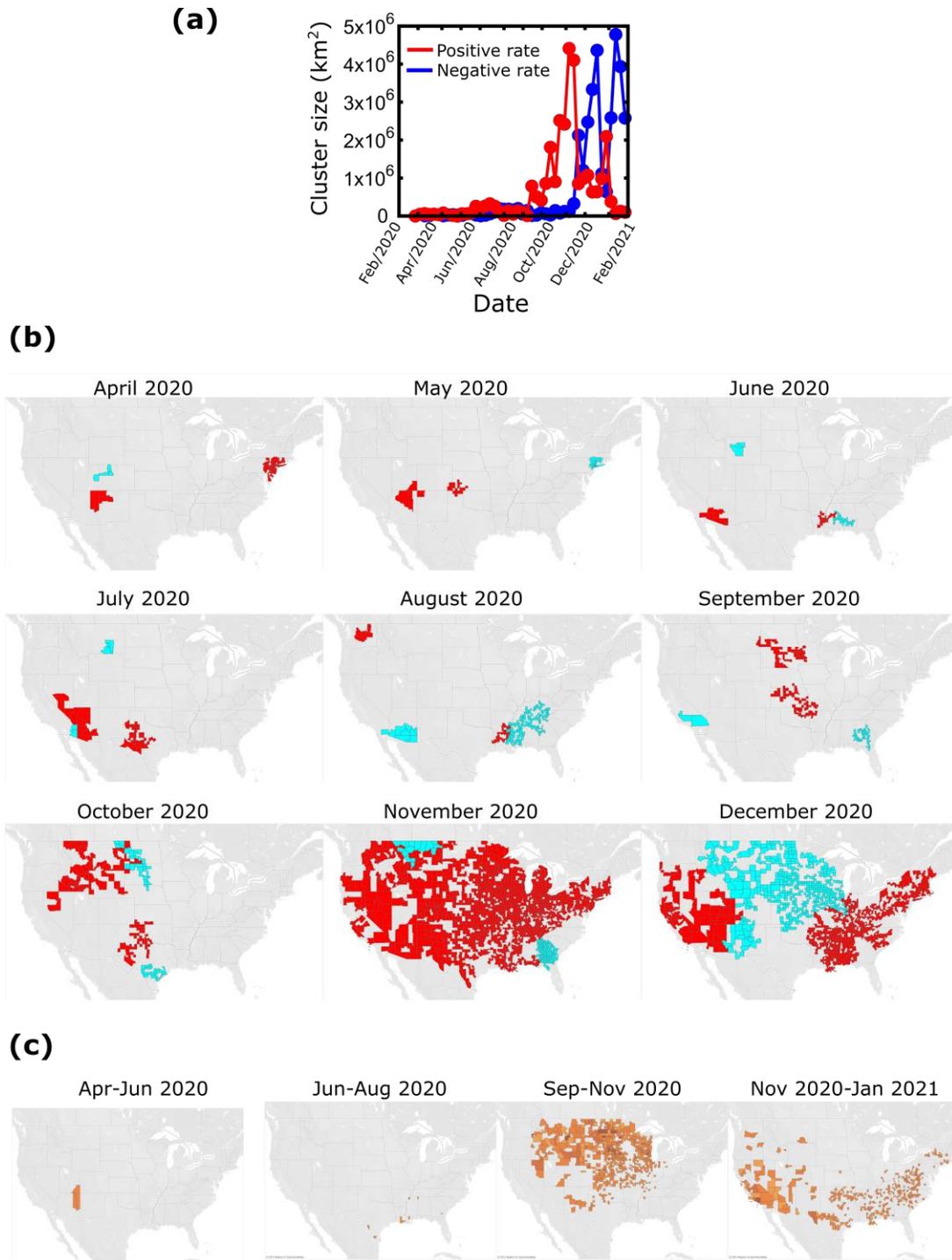

**FIGURE 5 | Clustering analysis of the correlations (a)** Total area of the largest clusters as a function of time. The red line corresponds to clusters of increasing number of cases and the blue line corresponds to clusters of decreasing cases. **(b)** The maps show the two largest clusters with positive rate (red) and the two largest clusters of negative rate (blue) at different times. **(c)** 'Persistence' maps. These maps show the counties which remained in positive clusters for more than half of the time period indicated on the map.